\documentclass[preprint,showpacs,preprintnumbers,amsmath,amssymb]{revtex4}

\usepackage{graphicx}
\usepackage{dcolumn}
\usepackage{bm}
\usepackage{epsfig}

\newcommand{\z}{&&\hspace*{-1cm}}
\newcommand{\ep}{\varepsilon}
\newcommand{\bea}{\begin{eqnarray}}
\newcommand{\eea}{\end{eqnarray}}
\newcommand{\be}{\begin{equation}}
\newcommand{\ee}{\end{equation}}

\begin{document}

\boldmath
\title{Gluon evolution for
  the Berger-Block-Tan form of the  
structure function $F_2$}
\unboldmath

\author{
A.V. Kotikov
}
\affiliation{
 Laboratory of Theoretical  Physics of the Joint Institute for Nuclear Research, 
141980 Dubna, Russia}
  
\date{\today}

\begin{abstract}
  We present a nonlinear 
  modification of the evolution of
  the gluon density, obtained at small $x$
  from the Berger-Block-Tan form
  of the deep inelastic structure function $F_2$
  in the leading
order of perturbation theory.

\end{abstract}

\maketitle

In the snall-$x$ regime, nonperturbative effects were expected to play an important role.
However, as it has been
observed up to very low $Q^2 \sim 1$ GeV$^2$ values,
considered processes are
described reasonably well by perturbative OCD (pQCD) methods
(see, for example, \cite{CooperSarkar:1997jk}).
It should be noted, nonetheless, that at extremely low $x$, $x \to 0$, the pQCD evolution provides a rather singular
behavior of the parton distribution functions (PDFs)
(see e. g. Refs. \cite{Kotikov:1998qt,Kotikov:2012sm} and references therein), which strongly
violates the Froissard boundary \cite{Froissart:1961ux}.

In Ref. \cite{Berger:2007vf}
a new form of the deep inelastic lepton-hadron scattering (DIS) structure function (SF) $F_2(x,Q^2)$
was proposed.
It will be called below as the Berger-Block-Tan (BBT) structure function.
The SF  $F_2^{\rm BBT}(x,Q^2)$ 
leads to the low $x$ 
asymptotics of the (reducted) DIS cross-sections $\sim \ln^2 1/x$, which is in turn in an agreement with
the Froissard predictions
\cite{Froissart:1961ux}. This parametrization is
relevant in
investigations of ultra-high energy processes, such as scattering of cosmic neutrino off hadrons
(see \cite{Fiore:2004nt,Illarionov:2011wc}).

Following to our previous studies in \cite{Kotikov:1994vb} and \cite{Kotikov:1994jh}, recently
the gluon density $f_g(x,Q^2)$ and the longitudinal DIS SF $F_L(x,Q^2)$  in the BBT form have been
obtained in Refs. \cite{Chernikova:2016xwx,Kotikov:2017cld} and \cite{Kaptari:2018sxh} at small values of $x$, using 
the SF $F_2^{\rm BBT}(x,Q^2)$.
To do it, we proposed a violation of twist-two evolution of gluon density by a nonlinear term.
The purpose of the present letter
to show the exact form of the violation.
All the results will be done at the leading order (LO) of perturbation theory.

{\bf 1.}
The SF  $F_2$ is expressed through
the quark density $f_q(x,Q^2)$
(hereafter $f_a(x,Q^2)$ $(a=s,g)$ are usual PDFs multiplied by $x$)
as
\be
F_2(x,Q^2) = e f_q(x,Q^2),~~~ e=\frac{1}{f} \sum^f_{i=1} e_i^2 \equiv \frac{e_{2f}}{f} \, , 
\label{F2}
\ee
where $e$ is the average charge square
and $f$ is the number of flavors.

The famous DGLAP equations \cite{Gribov:1972ri}
relate the gluon and quark densities
as
 \be
\frac{d f_a(x,Q^2)}{dlnQ^2}  = -\frac{a_s(Q^2)}{2} \sum_{b=s,g} P^{(0)}_{ab}
(x) \otimes  f_b(x,Q^2),~~ f_1(x) \otimes f_2(x) \equiv \int^1_x \frac{dy}{y} 
 f_1(y) f_2\left(\frac{x}{y}\right) \, ,
\label{DGLAP} 
\ee
where the symbol $\otimes$ stands the  Mellin convolution,
\be
a_s(Q^2) = \frac{\alpha_s(Q^2)}{4\pi} = \frac{1}{\beta_0\ln(Q^2/\Lambda^2_{\rm LO})}
\label{as}
\ee
is the strong coupling constant
and 
$P^{(0)}_{ab} (x)$ (hereafter $(a,b=s,g)$) and  $\beta_0$ are the LO splitting functions and the first coefficient of QCD $\beta$-function.

Equations (\ref{F2}) and (\ref{DGLAP}) lead to the following realtions
\bea
\z    \frac{dF_2(x,Q^2)}{dlnQ^2}  = -\frac{a_s(Q^2)}{2} \biggl[
e 
P^{(0)}_{sg} (x) \otimes f_g(x,Q^2) + 
P^{(0)}_{ss} (x)  \otimes  
F_2(x,Q^2)  
 \biggr] ,
\label{dF2} \\
\z 
\frac{d f_g(x,Q^2)}{dlnQ^2}  = -\frac{a_s(Q^2)}{2}  \biggl[
P^{(0)}_{gg} (x) \otimes f_g(x,Q^2) + e^{-1} 
P^{(0)}_{gs} (x)  \otimes   
F_2(x,Q^2)\biggr]  + f_{2g}(x,Q^2)
\label{fg} \, ,
\eea
where we add to the gluon evolution (\ref{fg}) the nonlinear
term  (following to
Refs.\cite{Gribov:1984tu,Zhu:1998hg})
\be
f_{2g}(x,Q^2)=
S(x,Q^2) \,  \tilde{f}^2_g(y,Q^2),~~ \tilde{f}^2_g(y,Q^2)=
\int_x^1 \frac{dy}{y} f^2_g(y,Q^2)
\label{f2g} \, 
\ee
with some extra function $S(x,Q^2)$, which is the subject
of the present study.

The nonlinear term
modifies strongly the low $x$ asymptotics of the gluon density, which becomes to be ruther flat.
Such types of gluon density leads to less singular form of the high-energy asymptotics of cross-sections
\cite{Fiore:2004nt}.

{\bf 2.}
Following to the BBT form $F_2^{\rm BBT}(x,Q^2)$
(see Eq. (\ref{n9}) below),
the equation (\ref{dF2}) can be considered as a definition of the gluon density $f_g^{\rm BBT}(x,Q^2)$.
One can substantially
simplify the calculations if one considers Eqs. (\ref{dF2}) and (\ref{fg}) in the space of Mellin momenta, by taking
advantage of the fact the Mellin convolution
becomes merely a product of
individual Mellin momenta of the corresponding functions in the space of Mellin momenta.
So, we go to Mellin space
\footnote{
  The results in Ref. \cite{Kotikov:2017cld} have been obtained directly in $x$-space following to the approach
  in \cite{Kotikov:1993xe}.}, using the following Mellin moments
\bea
&&M_2(n,Q^2) = \int^1_0 dx x^{n-2} F_2(x,Q^2),~~~
M_g(n,Q^2) = \int^1_0 dx x^{n-2} f_g(x,Q^2)\, ,
\label{Mg}\\
&&M_{2g}(n,Q^2) = \int^1_0 dx x^{n-2} S(x,Q^2) f^2_g(x,Q^2),~~~
\gamma^{(0)}_{ab} (n) = \int^1_0 dx x^{n-2} 
P^{(0)}_{ab} (x) \, .
\label{gamma}
\end{eqnarray}

After the Mellin transform, the
equations (\ref{dF2}) and (\ref{fg}) are replaced by ones 
\bea
&&\frac{dM_2(n,Q^2)}{dlnQ^2}  = -\frac{a_s(Q^2)}{2} \biggl[
e \gamma^{(0)}_{sg} (n) M_g(n,Q^2) + \gamma^{(0)}_{ss} (n)  
M_2(n,Q^2)  
 \biggr] ,
\label{dM2}\\
&&\frac{dM_g(n,Q^2)}{dlnQ^2}  = -\frac{a_s(Q^2)}{2} \biggl[
\gamma^{(0)}_{gg} (x) M_g(x,Q^2) + e^{-1} 
\gamma^{(0)}_{gs} (x) M_2(x,Q^2)
\biggr] + M_{2g}(n,Q^2) 
,
\label{dMg} 
\eea
with the  LO anomalous dimensions $\gamma^{(0)}_{ab} (n)$ 
(hereafter ($a,b=s,g$)), which are
 \bea
&&\gamma^{(0)}_{sg} (n) = - \frac{4f (n^2+n+2)}{n(n+1)(n+2)},~~~\gamma^{(0)}_{gs} (n) = - \frac{16 (n^2+n+2)}{3(n-1)n(n+1)},\nonumber \\
&&
\gamma^{(0)}_{ss} (n) ~=~ \frac{32}{3} \biggl[\Psi(n+1)-\Psi(1)
  -\frac{3}{4} + \frac{1}{2n(n+1)}
  \biggr],  \nonumber \\
&&
\gamma^{(0)}_{gg} (n) ~=~ 24
\biggl[\Psi(n+1)-\Psi(1)
  -\frac{1}{(n-1)n} - \frac{1}{(n+1)(n+2)}\biggr] + 2\beta_0, 
\label{ana}
\eea
where $\Psi(n)$ is the Euler $\Psi$-function.

{\bf 3.}~~  Following to Ref. \cite{Block:2011xb}, we use the extended  BBT form of the
proton SF at $x < x_P=0.011$, where the combine H1 and ZEUS data 
\cite{Aaron:2009aa} taken into account:
\be
F_{2}^{\rm BBT}(x,Q^2) =
(1-x) \, \sum_{m=0}^2 A_m(Q^2) L^m,~~  A_0 = \frac{F_P}{1-x_P},~~ F_P=0.413 \pm 0.003
\label{n9}
\ee
with
\be
L= \ln \left[\frac{(1-x) x_P}{(1-x_P)x}\right],~~
A_m(Q^2) = \sum_{k=0}^2 a_{mk} \,
l^k,~~
m=(1,2),~~
l= \ln(Q^2)
\label{n9.0}
\ee
and
\bea
&& a_{10}
\cdot 10^{2} = -8.471 \pm 0.260,~~
a_{11}
\cdot 10^{2} = 4.190 \pm 0.156,~~
a_{12}
\cdot 10^{3} = -3.973 \pm 0.213, \nonumber \\
&& a_{20}
\cdot 10^{2} = 1.297 \pm 0.036,~~
a_{21}
\cdot 10^{4}=
2.473 \pm 0.246,~~
a_{22} 
\cdot 10^{3} =
1.642 \pm 0.055 \, .
\label{n10}
\eea

We note that gluon density and
coupling constant are strongly correlated when they are fitted from experimental data.
Following this fact, we
will try to find the modified gluon density $\hat{f}_g^{\rm BBT}(x,Q^2) =a_s(Q^2)f_g^{\rm BBT}(x,Q^2)$
in the
form
\footnote{We
  omit the factor $(1-x)$, which contributes into Eq. (\ref{n9}).
It is in agreement with the quark counting rules \cite{Matveev:1973ra}, where $f_s(x)/f_g(x) \sim (1-x)$ at $x \to 1$.}
\be
\hat{f}_g^{\rm BBT}(x,Q^2) = \frac{3}{4e_2} \, \sum_{m=0}^2 B_m(Q^2) L^m,~~~
B_m(Q^2) = \sum_{k=0}^2 b_{mk} \,  l^k \, .
\label{n9.1}
\ee

Performing the Mellin transforms (\ref{Mg}) and (\ref{gamma}), we have the following representations
for the
$M_k(n,Q^2)$ (hereafter $(k=2,g)$) at $n=1+\omega$ and  $\omega \to 0$
\bea
M_{2}^{\rm BBT}(n,Q^2) &=& \frac{x_p^{\omega}}{\omega}
\biggl[ A_0 + \frac{1}{\omega} A_1(Q^2) 
+ \frac{2}{\omega^2}  A_2(Q^2) 
+ O(x_p)
\biggr] \, ,
\label{M2BBT}\\
M_{g}^{\rm BBT}(n,Q^2) &=& \frac{3}{4e_2} \, \frac{x_p^{\omega}}{\omega}
\biggl[ B_0(Q^2) + \frac{1}{\omega} B_1(Q^2) 
+ \frac{2}{\omega^2} B_2(Q^2) 
+ O(x_p)
\biggr] \, , 
\label{MgBBT}
\eea
which are obtained using the
basic integrals
\be
\int^{x_p}_0 dx \, x^{n-2} (1-x) L^k(x) = k!\frac{x_p^{\omega}}{\omega^{k+1}} +  O(x_p^{\omega+1}) .
\label{Int}
\ee

{\bf 3a.}~~ From Eq. (\ref{dM2})  we have ($\hat{M}_g(n,Q^2)= a_s(Q^2) M_g(n,Q^2)$) 
\bea
\hat{M}_g(n,Q^2) = -\frac{2}{e \gamma^{(0)}_{sg} (n)}
\left[\frac{dM_2(n,Q^2)}{dlnQ^2} +\frac{a_s(Q^2)}{2}  \gamma^{(0)}_{ss} (n) M_2(n,Q^2)
  \right]  .
\label{Mg_res}
\eea

Equating the coefficients in front of $1/\omega$ singularities
in the both sides, one finds 
the results for $B_m(Q^2)$
\be
B_2 = A'_2, ~~
B_1 =  A'_1 + \frac{13}{6} A'_2  + \frac{32 a_s}{3}
\delta_{ss} A_2,~~
B_0  =
\frac{13}{12} A'_1 - \frac{1}{8} A'_2 +  \frac{16a_s}{3}
  \Bigl(\delta_{ss} A_1 + 2\Delta_{ss}
  A_2 \Bigr),
\label{9.7} 
\ee
where
\be
A'_m(Q^2) =
\frac{A_m(Q^2)}{dl} 
=   a_{m,1} +  2 a_{m,2} \, l,~~ (m=1,2),
\label{9.5}
\ee
and
\be
\delta_{ss}=\frac{5}{2}-\zeta_2,~~ \Delta_{ss}=\frac{13}{12}\zeta_2 -\zeta_3-\frac{11}{96}
\label{DeltasSS}
\ee 
are $\omega$ and  $\omega^2$ terms of the $\omega$-expansion of $ \gamma^{(0)}_{ss}(1+\omega)$.
We used also the propoerty $ \gamma^{(0)}_{ss}(1)=0$.

{\bf 3b.}~~ Now we consider the equation (\ref{dMg}) for the evolution of the gluon density, which contains
the Mellin moment $M_{2g}(n,Q^2)$
of the nonlinear term $f_{2g}(x,Q^2)=S(x,Q^2) \tilde{f}^2_{g}(x,Q^2)= \hat{S}(x,Q^2) \hat{f}^2_{g}(x,Q^2)$ with $\hat{f}^2_{g}(x,Q^2)=
a^2_s(Q^2) \tilde{f}^2_{g}(x,Q^2)$
and $\hat{S}(x,Q^2)=S(x,Q^2)/a_s^2(Q^2)$:
\be
M_{2g}(n,Q^2)  =  \frac{1}{2} \biggl[\gamma^{(0)}_{gg} (n) \hat{M}_g(n,Q^2) + e^{-1} \gamma^{(0)}_{gs} (n) a_s(Q^2)
  \, M_2(n,Q^2)
\biggr] - \frac{d}{dlnQ^2} \, \frac{\hat{M}_g(n,Q^2)}{a_s(Q^2)},
\label{M2g} 
\ee
where all terms in the r.h.s. are already fixed.

We note that the anomalous dimensions $\gamma^{(0)}_{gs} (n)$ and $\gamma^{(0)}_{gg} (n)$ contain singularities at the
limit $n \to 1$
(see
eq. (\ref{ana})), which produce the asymptotics $\sim L^3$
in the first two terms of the r.h.s.
  of (\ref{fg}) and, correspondly, the poles $\sim 1/\omega^4$ in the first two terms of the r.h.s.
  of (\ref{dMg}).
  The function $\hat{f}^2_{g}(x,Q^2)$ has the asymptotics $\sim L^5$
  \be
 \hat{f}^2_{g}(x,Q^2) =\frac{9L^3}{16 e_2} \, \sum_{m=0}^2 D_m(Q^2) L^{m} + O(L^2) \, ,
\label{hat2g}
\ee
where
\be
D_0= \frac{1}{3} \, \Bigl(B_1^2 + 2 B_0B_2\Bigr),~~D_1= \frac{1}{2} \, B_1 B_2,~~ ~D_2= \frac{1}{5} \, B_2^2 \, .
\label{Di}
\ee
and, thus,
from self-consistence of the small $x$-asymptotics, the coefficient $\hat{S}(x,Q^2)$ should be like $1/L^{2}$.
So, we can use
the function $\hat{S}(x,Q^2)$
in the following form
  \be
  \hat{S}(x,Q^2) =\sum_{m=0}^2 \hat{S}_m(Q^2) \, L^{m-4} \, .
\label{Rpara}
\ee

Then, the last term in the  r.h.s.
of (\ref{fg}) can be parameterized as
\be
 f_{2g}(x,Q^2) =\frac{9L}{16 e_2} \, D_2(Q^2) \sum_{m=0}^2 \tilde{S}_m(Q^2) L^{m} + O(L^0) \, ,
\label{f2gPara}
\ee
where $\tilde{S}_2=  \hat{S}_2$ and
\be
\tilde{S}_1 =\hat{S}_1  + \frac{D_1}{D_2} \hat{S}_0,~~
\tilde{S}_0 =\hat{S}_0  + \frac{D_1}{D_2} \hat{S}_1 + \frac{D_0}{D_2}  \hat{S}_ .
\label{tR}
\ee

The Mellin moment of
$f_{2g}(x,Q^2)$  in  (\ref{f2gPara}) is 
\be
M_{2g}(n,Q^2) = \frac{x^p}{\omega} \, \frac{9 D_2}{16 e_2^2} \, \left( \tilde{S}_0 \, \frac{1}{\omega} + \tilde{S}_1 \frac{2}{\omega^2}
+ \tilde{S}_2  \frac{6}{\omega^3}  \right) + O\left(\frac{1}{\omega}\right)   \, 
\label{M2gPara}
\ee
and obeys the relation (\ref{M2g}).
Comparing the coefficients in front of $1/\omega$ singularities
in the l.h.s and r.h.s. in (\ref{M2g}) and using relations (\ref{Di}), we have for the coefficients
$\tilde{S}_m(Q^2)$
\bea
\z   \tilde{S}_2 = \hat{S}_2  = -\frac{80 e_2}{3 B^2_2} \, \left[B_2 + \frac{8f}{27} a_s A_2\right],
\nonumber \\
\z  \tilde{S}_1 = \frac{20 e_2}{3} \, \frac{\overline{B}_2}{ B^2_2}  - \frac{40 e_2 }{B^2_2} \, \left[B_1-2\ep_{gg} B_2 +
  \frac{8f}{7} a_s
 \left(A_1- \frac{3}{2} A_2\right)\right], \nonumber \\
\z  \tilde{S}_0 = \frac{20 e_2}{3} \, \frac{\overline{B}_1}{ B^2_2}  - \frac{80 e_2}{B^2_2 } \, \Biggl[B_0-\ep_{gg} B_1-2\delta_{gg} B_2
  +
  \frac{8f}{27} a_s
 \left(A_0- \frac{3}{4}A_1+\frac{7}{4}A_2 \right)\Biggr],
\label{tR.1} 
\eea
where
\be
\overline{B}_m(Q^2) = \frac{d}{dl} \, \left(\frac{B_m(Q^2)}{a_s(Q^2)}\right) \, .
\label{tB} 
\ee

Inverting the relations (\ref{tR}) as
\be
\hat{S}_1=\tilde{S}_1 - \frac{5B_1}{2B_2} \tilde{S}_0,~~
\hat{S}_0=\tilde{S}_0 - \frac{5B_1}{2B_2} \tilde{S}_1 + \left(\frac{55}{12} \, \frac{B^2_1}{B^2_2}- \frac{10}{3} \,
\frac{B_1}{B_2}\right) \tilde{S}_0 
\label{hR}
\ee
and taking Eq.
(\ref{tR.1}), after some algebra  we have for the coefficients $\hat{S}_m(Q^2)$
\bea
\z  \hat{S}_1= \frac{20 e_2}{3} \, \frac{\overline{B}_2}{B^2_2}  + \frac{40 e_2}{B^2_2} \, \Biggl[\frac{2}{3} \, B_1+2\ep_{gg} B_2
  - \frac{8f}{27} a_s
 \left(A_1- \left[\frac{3}{2} + \frac{5B_1}{3B_2} \right] A_2\right)\Biggr], \nonumber \\
\z  \hat{S}_0= \frac{10 e_2}{3B^3_2} \, \Bigl(2\overline{B}_1 B_2 - 5 B_1 \overline{B}_2\Bigr)
  - \frac{80 e_2}{B^2_2(Q^2)} \, \Biggl[B_0-\left(\frac{10}{9}- \frac{3}{2}\ep_{gg}\right) B_1 + \frac{5}{18}  \, \frac{\overline{B}^2_1}{B_2}
    -2\delta_{gg} B_2 \nonumber \\
  &&
  \hspace{2cm}  +
  \frac{8f}{27} a_s
  \left(A_0- \left[\frac{3}{4} + \frac{5B_1}{4B_2} \right] A_1+ \left[\frac{7}{4} + \frac{55}{72} \, \frac{B_1}{B_2}
    +  \frac{55}{36} \, \frac{B_1^2}{B_2^2}  \right] A_2 \right)\Biggr] .
\label{hR.1} 
\eea

So, now the coefficient function $\hat{S}(x,Q^2)$
in (\ref{f2g}) is completely
fixed by the
parameters (\ref{n9.0}) of the SF $F_2^{\rm BBT}(x,Q^2)$ in  (\ref{n9}).

{\bf Conclusion.}
Using
the SF $F_2^{\rm BBT}(x,Q^2)$ and
the gluon density $f_g^{\rm BBT}(x,Q^2)$, we recovered
the form of the coefficient function $\hat{S}(x,Q^2)$ in front of the nonlinear
term in the gluon evolution
at the LO
of perturbation theory.
The coefficient function is negative at low $x$ values and is
suppressed  as $1/\ln^2(1/x)$ (see eq. (\ref{Rpara})),
that is in a full agreement with earlier studies in 
\cite{Levin:1990gg}. However, the $Q^2$-dependence is different: at large $Q^2$ values we have $\hat{S}(x,Q^2) \sim 1/ln(Q^2)$ but the
corresponding coefficient function in Ref. \cite{Levin:1990gg} $\sim 1/Q^2$. 
Our $Q^2$-dependence comes directly from the shape (\ref{n9}) of the SF $F_2^{\rm BBT}(x,Q^2)$, where there is only $ln(Q^2)$ dependence of
the coefficients (\ref{n9.0}).

As it was already discussed,
the nonlinear term in the gluon evolution (\ref{f2g})
leads to  significantly less singular form (\ref{n9.1})
of the gluon density $f_g^{\rm BBT}(x,Q^2)$ 
in comparison with the corresponding results in perturbation theory (see, for
example, \cite{Kotikov:1998qt}). Such form is in an agreement with
the Froissard boundary \cite{Froissart:1961ux} and 
can be useful for studying \cite{Fiore:2004nt,Illarionov:2011wc,Bertone:2018dse}
the high-energy behavior of photon, neutrino and proton cross sections.

\end{document}